\documentclass[authorversion, manuscript]{acmart}

\makeatletter
\renewcommand\@formatdoi[1]{\ignorespaces}
\makeatother
\renewcommand\footnotetextcopyrightpermission[1]{}

\AtBeginDocument{%
  }

\copyrightyear{2024}
\acmYear{2024}
\setcopyright{rightsretained}

\acmConference[CSCW'24]{The 27th ACM SIGCHI Conference on Computer-Supported Cooperative Work \& Social Computing: Collective Imaginaries for the Futures of Care Work Workshop}{November 9, 2024}{Costa Rica, US}
\acmBooktitle{CSCW'24: The 27th ACM SIGCHI Conference on Computer-Supported Cooperative Work \& Social Computing: Collective Imaginaries for the Futures of Care Work Workshop, November 9, 2024, Costa Rica, US}

\begin{document}
    
    \title{Imagining Better AI-Enabled Healthcare Futures: The Case for Care By Design}

    \author{Kevin Doherty}
    \email{kevin.doherty@ucd.ie}
    \orcid{0000-0002-7647-5139}
    \affiliation{%
      \institution{School of Information and Communication Studies, University College Dublin}
      \city{Dublin}
      \country{Ireland}
    }

    \author{Emma Kallina}
    \email{emk45@cam.ac.uk}
    \orcid{0000-0003-4912-7216}
    \affiliation{%
      \institution{University of Cambridge}
      \city{Cambridge}
      \country{United Kingdom}
    }

    \author{Kayley Moylan}
    \email{kayley.moylan@ucdconnect.ie}
    \orcid{0009-0001-4184-6913}
    \affiliation{%
      \institution{School of Information and Communication Studies, University College Dublin}
      \city{Dublin}
      \country{Ireland}
    }

    \author{María Paula Silva}
    \email{maria.p.silva@ucdconnect.ie}
    \orcid{0009-0007-8665-2438}
    \affiliation{%
      \institution{School of Information and Communication Studies, University College Dublin}
      \city{Dublin}
      \country{Ireland}
    }

    \author{Sajjad Karimian}
    \email{sajjad.karimian@ucdconnect.ie}
    \orcid{0000-0002-9772-144X}
    \affiliation{%
      \institution{School of Computer Science, University College Dublin}
      \city{Dublin}
      \country{Ireland}
    }

    \author{Shivam Shumsher}
    \email{shivam.shumsher@ucdconnect.ie}
    \orcid{0009-0008-4078-4969}
    \affiliation{%
      \institution{School of Information and Communication Studies, University College Dublin}
      \city{Dublin}
      \country{Ireland}
    }

   \author{Rob Brennan}
    \email{rob.brennan@ucd.ie}
    \orcid{0000-0001-8236-362X}
    \affiliation{%
      \institution{School of Computer Science, University College Dublin}
      \city{Dublin}
      \country{Ireland}
    }
    
    \renewcommand{\shortauthors}{Doherty et al.}

    \begin{abstract}
    
        We find ourselves on the ever-shifting cusp of an AI revolution --- with potentially metamorphic implications for the future practice of healthcare. For many, such innovations cannot come quickly enough; as healthcare systems worldwide struggle to keep up with the ever-changing needs of our populations. And yet, the potential of AI tools and systems to shape healthcare is as often approached with great trepidation as celebrated by health professionals and patients alike. These fears alight not only in the form of privacy and security concerns but for the potential of AI tools to reduce patients to datapoints and professionals to aggregators --- to make healthcare, in short, less caring. This infixated concern, we - as designers, developers and researchers of AI systems - believe it essential we tackle head on; if we are not only to overcome the AI implementation gap, but realise the potential of AI systems to truly augment human-centred practices of care. This, we argue we might yet achieve by realising newly-accessible practices of AI healthcare innovation, engaging providers, recipients and affected communities of care in the inclusive design of AI tools we may yet enthusiastically embrace to materialise new opportunities for increasingly meaningful and effective care.    
      
    \end{abstract}

    
    
    \received{23 October 2024}
    \received[revised]{23 October 2024}
    \received[accepted]{23 October 2024}
    
\maketitle

\section{The Case for \textit{AI Health}Care}

    Healthcare systems worldwide are under ever-growing pressures, often described a result of ageing populations and growing rates of chronic, co-morbid medical conditions as much as the under-funding of services \cite{andersen2019challenging, world2005mental}. Innovative solutions to this long-standing and ever anew challenging situation are required, we are told by policy-makers, health system managers, researchers and developers alike. For many years, digital technologies from electronic patient records to shared-decision making tools have been promoted as the answer --- a discourse only further accented today by the recent arrival of AI technologies we are told will ``help save lives by transforming healthcare'' through ``more personalized, accessible and effective solutions'' \cite{google2024}. Indeed, AI systems offer great potential to address many urgent healthcare challenges, improve patient outcomes, and advance the adoption of digital tools to support healthcare \cite{matheny2020artificial}.

    And yet, despite such promise, many doctors speak to us still today of the degree to which they ``hate their computers'' \cite{gawande2018doctors}. This claim, as made by surgeon Atul Gawande in an influential New Yorker Annals of Medicine article (ibid.), is equally a refrain heard often by those of us engaged in the frontline practice of the design, development and evaluation of novel digital health solutions. Despite billions spent and decades of effort, still today the vast majority of digital clinical innovations either fail outright or are deployed only to be met with practices of passive hindrance and resistive compliance if not active resistance by health professionals \cite{ziebland2021power, zajkac2023clinician} --- each often visions of AI in-action which yet fail to function in-practice \cite{aristidou2022bridging, coiera2019last}. And the primary cause of this state of affairs, the weight of the human-computer interaction (HCI) and computer-supported cooperative work (CSCW) literature suggests, remains a lack of consideration for human factors; including clinicians' needs, values, and ways of working \cite{yang2019unremarkable, zajkac2023clinician, shaw2019artificial, andersen2021realizing, jacobs2021designing}.
 
    This lack of consideration, we see increasingly realised in the form of practices of resistance to the perceived imposition of AI systems, even those with seemingly-impressive evidence for their capacity to improve care \cite{yang2019unremarkable}. At the root of these fears, we often hear, lie concerns about datafication and inaccuracies \cite{schull2016data}, exacerbated healthcare disparities \cite{matheny2020artificial, obermeyer2019racial-bias}, confused lines of responsibility \cite{doherty2020personal} and increased workloads \cite{yang2019unremarkable}. Such responses might then yet give us pause to ask; is this opposition to the technology, or to the assumptions about care we promote, embed and occasionally even enforce therein? In other words, \textbf{are we tackling the right problems}? If not approached with care, the decisions we make in the design of systems developed to address challenges in the practice of care, may instead worsen the underlying causes of crisis.

    \subsection{Future Possible \textit{AI-Enabled Healthcare} Crises} \label{sec:issues}
    
        \noindent\hrulefill
        
        \noindent Indeed, should we in the development of AI-enabled systems of care inadvertently reify existing problematic assumptions, we may come to create a healthcare future in which clinicians find themselves employing systems they loathe \cite{ziebland2021power} for purposes they feel reduce their patients to potentially misleading data points \cite{schull2016data} and their professional roles to automatons \cite{gawande2018doctors} --- such that AI serves less as a meaningful lever for care than a ``crowbar to pry dignity out of our collective hands'' \cite{geuter2024}. 
   
        \noindent\hrulefill
    
    Much work has yet been conducted to date to avoid just such scenarios --- researchers' own efforts towards not only resistance but awareness-raising. This includes calls for consideration of bias, trust and transparency in the research and development of AI systems \cite{waefler2020explainability, kim2023organizational}, as encapsulated in the transparency requirements of the European Union's (EU) AI Act, and new EU Ethics Guidelines for Trustworthy AI requiring developers to support ``diversity, non-discrimination and fairness'' for the benefit of ``all human beings'' \cite{ai2019high, van2024eu}. These commendable efforts yet surface both the desire for action and the immense challenges of bridging AI theory, policy and the realities of AI systems development --- MIT's recently-developed AI-Risk Repository highlighting more than 700 AI risks categorized by cause and risk domain \cite{slattery2024ai}.

    HCI and CSCW researchers and practitioners have at the same time long advocated, as one path towards the design of more desirable AI-enabled futures of care, for the involvement of users in the design of digital tools and systems \cite{jacobs2021designing, andersen2021realizing, karusala2021future}. Drawing on human factors ergonomics, design-thinking, user-centred and service-design traditions, we as a collective community have developed and employed numerous methods to promote and enable greater participation in design and research activities spanning the product-development lifecycle by a wide range of stakeholders; from clinicians to patients and diverse other caregivers. HCI and Design as academic and professional practices have therefore today granted us a diverse portfolio of tools and methodologies for the user-centred design of digital tools and systems — from ideation, brainstorming and user-centred evaluation techniques to creative methods of low- and high-fidelity prototyping \cite{chivukula2024surveying, hsieh2023what, CARAYON2020103033}. And yet, we at the same time recognise the design of Trustworthy AI (TAI) tools as presenting new and significant challenges to the adoption and adaptation of these practices \cite{sadek2024challenges}.

    Indeed, a number of researchers have turned their focus in recent years towards figuring out just how such methods may be adapted to permit stakeholder-engagement in the design of AI systems in particular --- producing notable conceptual tools from Carnegie Mellon's AI Brainstorming \cite{yildirim2023creating} to Microsoft's Human-AI eXperience (HAX) toolkits \cite{vorvoreanu2023create} (for more see \cite{datacentre_toolslibrary}). While other researchers have even begun to adopt insider-ethnographer and meta-research perspectives, seeking to uncover problems and processes for the more effective development of tools - conceptual and tangible - to realise more sustainable and responsible AI futures \cite{kawakami2024situate, wong2023seeing, sadek2024guidelines}.

    Healthcare is yet a high-stakes context imposing specific demands on AI systems. In this domain, it is essential that AI systems not only prove accurate, reliable and avoid exacerbating inequalities, but refrain from dehumanising users, detracting from the meaningfulness of clinical work, nor devaluing the very heart of this most human of endeavours --- care itself \cite{waefler2020explainability, doherty2022unboxing, doi:10.1080/07853890.2024.2302980}. These are fears motivated in many respects by shared concern for the forces of rationalisation and determinism --- whether wrought by perverse commercial incentives, the reductionist tendencies of technology or a neo-liberal politics of austerity. And the complex, hyped and often opaque nature of AI systems does yet increasingly risk excluding key stakeholders from essential conversations about the present and future direction of these technologies. It is, in this socio-technical context, unsurprising to see HCI and CSCW's advocacy for \textbf{human-centred design} practices increasingly mirrored by ever-more ardent calls among medical practitioners and researchers alike for a renewed focus on \textbf{patient-centred care}. Connecting these threads, strikes us, authors and professionals spanning diverse disciplines, as increasingly critical given \textbf{it is then not only technology but healthcare itself which many have argued requires greater care} \cite{montori2020we}. It is in the care of healthcare that patients and professionals alike find meaning. And care then challenges us to think more deeply about what we perceive technology as `for'. We are yet however to quite articulate in theory nor practice, just \textbf{how AI systems might be designed to support care as a fundamental component of the practice of healthcare}. 

\section{Towards Inclusive \textit{AI-Enabled Healthcare} Futures: The Case for Care}

    Which healthcare futures might we then seek to meaningfully enable through the design and adoption of AI systems? \textbf{Our position is that we must urgently recentre the care in healthcare; and that AI can provide us the moment to do just this}. This will yet require a complete re-imagining of just what is possible through technology --- design for efficacious, professional practices \textit{and} meaningful human experiences before incrementalism and datafication. We might indeed imagine AI systems deployed in clinical practice to make care not only more scalable, efficient and effective but to create new opportunities to materialise care in increasingly human-centred ways. To get there, the irony may be that we must first focus on helping healthcare to become more caring --- without insisting on AI as the means to do so.
    
    Arriving at such a future will yet urgently require the development and advancement of newly inclusive and accessible practices of AI healthcare innovation --- centering healthcare professionals, patients and caregivers; each members of a wider care network. Undertaking such work with its sustainability in mind requires the building of meaningful teams, confidence and capacities across disciplines and professions. And this task we have embraced this Autumn in Dublin, Ireland, establishing a new \textit{Inclusive Design for AI Healthcare Innovation (ID4AI) Network}, bringing together diverse medical, computer science and design professionals with the collective ambition of realising increasingly inclusive cultures and responsible practices of AI healthcare innovation.\footnote{Members of the ID4AI Network currently span the Schools of Information and Communication Studies and Computer Science at University College Dublin (UCD), the UCD Innovation Academy, the National College of Art and Design, St. James's Hospital, and the HSE Spark Innovation Programme.} Through this work, we attend to the following themes;

    \textbf{Embracing Ecologies of Care Work.} AI systems deployed in healthcare risk fragmenting care, augmenting disillusionment within teams, and creating greater distance between patients and professionals. If we are yet able to engage interdisciplinary medical teams in design process without reinforcing adverse power dynamics, attend to design as a practice best conducted in-situ rather than divorced from the ways in which every unique context is best placed to materialise care, and hold at the same time complementary models of medical and social care without reifying either one --- AI may just yet come to form a meaningful component of a portfolio of tools for care.
    
    \textbf{Enabling Ethical Care Work Practices.} AI healthcare systems also risk introducing bias, reducing decision-making to considerations of productivity, and rendering care less sensitive to individual difference. Ethical design is yet inclusive design, and inclusive design, effective design. By defining the desirable futures we work towards through patients' and health professionals' perspectives, creating sustainable infrastructure for caring innovation, and realising the active participation of clinical professionals in the inclusive co-design of AI systems, we aspire to develop AI tools enthusiastically adopted in favour of fairer systems of care \cite{nunes2022scoping}.

    \textbf{Ensuring Job Sustainability.} AI risks reducing the felt meaning of the professional practice of healthcare; as much as the rise and expansion of AI in healthcare risks making the design of digital systems itself a less inclusive practice, undoing years of progress towards increasingly person- and patient-centred practices of care. This, unless designed to enable appropriate practices of boundary-setting in support of clinicians' comfort, support wellbeing through sustainable work practices, and ultimately create time for richer connections and more caring conversations.

    \textbf{Co-Designing with Care Workers.} AI, however, also, risks providing carers even less of a voice. And the co-design of AI healthcare systems must yet create space not only to mirror the status quo of care but to challenge existing practices; including the valuing of rationalistic over humanistic forms of care. If we wish for patients, clinicians and communities of care to come to engage enthusiastically with AI healthcare systems, we must provide these diverse stakeholders the means and opportunities to engage in the co-creation of the tools which will increasingly come to shape our health. The solution, we suspect, must be twofold, encompassing both stakeholder training - educational practices to permit equal participation in design and ideation through effective articulation of the functioning and potential of AI tools - and a toolkit of novel means of imagining, communicating, representing and prototyping AI tools, as users' experiences with them, for the healthcare context. These challenges overcome, AI has the potential to be championed by rather than imposed upon care workers. And, \textbf{we therefore see this most recent `AI revolution' as an inflection point; as a moment of change, an opportunity to offer new more compassionate pathways and novel perspectives on care rather than to replicate existing problems, assumptions and inadequacies}.

    Our ambitions for this network are then to lead a programme of mixed-methods research yielding a toolkit for inclusive, sustainable practices of innovation, to grant practising health professionals the skills and experience to participate in AI healthcare ideation, and to build a network to advance the responsible, participatory design of TAI systems for real-world healthcare innovation, in part by redefining the role of technical team members as enablers of reflection among care communities; towards identifying opportunities for better AI-enabled futures through collective dialogue. \textbf{Our belief; that to build AI systems for more caring futures we require designers enthusiastic yet sceptical; half-sure \textit{and} whole-hearted}.

\begin{acks}
    
    We would like to acknowledge all other current members of the \textit{Inclusive Design for AI Healthcare Innovation Network} for your commitment and contributions towards our collective best possible AI-enabled healthcare futures; Dr. Emma Creighton and Enda O'Dowd of the National College of Art and Design Dublin, Dermot Burke of the Health Service Executive (HSE) Spark Innovation Programme, Dr. Marie E. Ward of St James's Hospital Dublin, and Jiaqi Zhang and William Davis of UCD's Innovation Academy.
    
\end{acks}
    
\bibliographystyle{ACM-Reference-Format}
\bibliography{bibliography}


\begin{thebibliography}{37}


\ifx \showCODEN    \undefined \def \showCODEN     #1{\unskip}     \fi
\ifx \showDOI      \undefined \def \showDOI       #1{#1}\fi
\ifx \showISBNx    \undefined \def \showISBNx     #1{\unskip}     \fi
\ifx \showISBNxiii \undefined \def \showISBNxiii  #1{\unskip}     \fi
\ifx \showISSN     \undefined \def \showISSN      #1{\unskip}     \fi
\ifx \showLCCN     \undefined \def \showLCCN      #1{\unskip}     \fi
\ifx \shownote     \undefined \def \shownote      #1{#1}          \fi
\ifx \showarticletitle \undefined \def \showarticletitle #1{#1}   \fi
\ifx \showURL      \undefined \def \showURL       {\relax}        \fi
\providecommand\bibfield[2]{#2}
\providecommand\bibinfo[2]{#2}
\providecommand\natexlab[1]{#1}
\providecommand\showeprint[2][]{arXiv:#2}

\bibitem[AI(2019)]%
        {ai2019high}
\bibfield{author}{\bibinfo{person}{HLEG AI}.} \bibinfo{year}{2019}\natexlab{}.
\newblock \showarticletitle{High-level expert group on artificial intelligence}.
\newblock \bibinfo{journal}{\emph{Ethics guidelines for trustworthy AI}}  \bibinfo{volume}{6} (\bibinfo{year}{2019}).
\newblock


\bibitem[Andersen et~al\mbox{.}(2019)]%
        {andersen2019challenging}
\bibfield{author}{\bibinfo{person}{Julie~H{\o}gsgaard Andersen}, \bibinfo{person}{Tine Tj{\o}rnh{\o}j-Thomsen}, \bibinfo{person}{Susanne Reventlow}, {and} \bibinfo{person}{Annette~Sofie Davidsen}.} \bibinfo{year}{2019}\natexlab{}.
\newblock \showarticletitle{Challenging Care Work: General Practitioners’ Perspectives on Caring for Young Adults With Complex Psychosocial Problems}.
\newblock \bibinfo{journal}{\emph{Health}} (\bibinfo{year}{2019}), \bibinfo{pages}{1363459319874100}.
\newblock


\bibitem[Aristidou et~al\mbox{.}(2022)]%
        {aristidou2022bridging}
\bibfield{author}{\bibinfo{person}{Angela Aristidou}, \bibinfo{person}{Rajesh Jena}, {and} \bibinfo{person}{Eric~J Topol}.} \bibinfo{year}{2022}\natexlab{}.
\newblock \showarticletitle{Bridging the chasm between AI and clinical implementation}.
\newblock \bibinfo{journal}{\emph{The Lancet}} \bibinfo{volume}{399}, \bibinfo{number}{10325} (\bibinfo{year}{2022}), \bibinfo{pages}{620}.
\newblock


\bibitem[Carayon et~al\mbox{.}(2020)]%
        {CARAYON2020103033}
\bibfield{author}{\bibinfo{person}{Pascale Carayon}, \bibinfo{person}{Abigail Wooldridge}, \bibinfo{person}{Peter Hoonakker}, \bibinfo{person}{Ann~Schoofs Hundt}, {and} \bibinfo{person}{Michelle~M. Kelly}.} \bibinfo{year}{2020}\natexlab{}.
\newblock \showarticletitle{SEIPS 3.0: Human-centered design of the patient journey for patient safety}.
\newblock \bibinfo{journal}{\emph{Applied Ergonomics}}  \bibinfo{volume}{84} (\bibinfo{year}{2020}), \bibinfo{pages}{103033}.
\newblock
\showISSN{0003-6870}
\urldef\tempurl%
\url{https://doi.org/10.1016/j.apergo.2019.103033}
\showDOI{\tempurl}


\bibitem[Chivukula et~al\mbox{.}(2024)]%
        {chivukula2024surveying}
\bibfield{author}{\bibinfo{person}{Shruthi~Sai Chivukula}, \bibinfo{person}{Colin Gray}, \bibinfo{person}{Ziqing Li}, \bibinfo{person}{Anne~C. Pivonka}, {and} \bibinfo{person}{Jingning Chen}.} \bibinfo{year}{2024}\natexlab{}.
\newblock \showarticletitle{Surveying a Landscape of Ethics-Focused Design Methods}.
\newblock \bibinfo{journal}{\emph{ACM J. Responsib. Comput.}} \bibinfo{volume}{1}, \bibinfo{number}{3}, Article \bibinfo{articleno}{22} (\bibinfo{date}{Sept.} \bibinfo{year}{2024}), \bibinfo{numpages}{32}~pages.
\newblock
\urldef\tempurl%
\url{https://doi.org/10.1145/3678988}
\showDOI{\tempurl}


\bibitem[Coiera(2019)]%
        {coiera2019last}
\bibfield{author}{\bibinfo{person}{Enrico Coiera}.} \bibinfo{year}{2019}\natexlab{}.
\newblock \showarticletitle{The last mile: where artificial intelligence meets reality}.
\newblock \bibinfo{journal}{\emph{Journal of medical Internet research}} \bibinfo{volume}{21}, \bibinfo{number}{11} (\bibinfo{year}{2019}), \bibinfo{pages}{16323}.
\newblock


\bibitem[Data and Society({[n.\,d.]})]%
        {datacentre_toolslibrary}
\bibfield{author}{\bibinfo{person}{Knowledge~Centre Data} {and} \bibinfo{person}{Society}.} \bibinfo{year}{[n.\,d.]}\natexlab{}.
\newblock \bibinfo{title}{Tool Library}.
\newblock
\newblock
\urldef\tempurl%
\url{https://data-en-maatschappij.ai/en/tools}
\showURL{%
\tempurl}
\newblock
\shownote{(accessed on 22.10.2024)}.


\bibitem[Doherty et~al\mbox{.}(2022)]%
        {doherty2022unboxing}
\bibfield{author}{\bibinfo{person}{Kevin Doherty}, \bibinfo{person}{Per B{\ae}kgaard}, \bibinfo{person}{Maria~Haahr Nielsen}, \bibinfo{person}{Alexandra Brandt~Ryborg J{\o}nsson}, \bibinfo{person}{Susanne Reventlow}, {and} \bibinfo{person}{Jakob~E Bardram}.} \bibinfo{year}{2022}\natexlab{}.
\newblock \showarticletitle{Unboxing the Clinical Health Technology Deployment}.
\newblock \bibinfo{journal}{\emph{IEEE Pervasive Computing}} \bibinfo{volume}{21}, \bibinfo{number}{4} (\bibinfo{year}{2022}), \bibinfo{pages}{64--73}.
\newblock


\bibitem[Doherty et~al\mbox{.}(2020)]%
        {doherty2020personal}
\bibfield{author}{\bibinfo{person}{Kevin Doherty}, \bibinfo{person}{Marguerite Barry}, \bibinfo{person}{Jos{\'e}~Marcano Belisario}, \bibinfo{person}{Cecily Morrison}, \bibinfo{person}{Josip Car}, {and} \bibinfo{person}{Gavin Doherty}.} \bibinfo{year}{2020}\natexlab{}.
\newblock \showarticletitle{Personal information and public health: Design tensions in sharing and monitoring wellbeing in pregnancy}.
\newblock \bibinfo{journal}{\emph{International journal of human-computer studies}}  \bibinfo{volume}{135} (\bibinfo{year}{2020}), \bibinfo{pages}{102373}.
\newblock


\bibitem[Gawande(2018)]%
        {gawande2018doctors}
\bibfield{author}{\bibinfo{person}{Atul Gawande}.} \bibinfo{year}{2018}\natexlab{}.
\newblock \showarticletitle{Why Doctors Hate Their Computers}.
\newblock \bibinfo{journal}{\emph{The New Yorker}}  \bibinfo{volume}{November 12} (\bibinfo{year}{2018}).
\newblock


\bibitem[Geuter(2024)]%
        {geuter2024}
\bibfield{author}{\bibinfo{person}{Jürgen Geuter}.} \bibinfo{year}{2024}\natexlab{}.
\newblock \bibinfo{title}{{A Choice}}.
\newblock \bibinfo{howpublished}{\url{https://tante.cc/2024/09/24/a-choice}}.
\newblock
\newblock
\shownote{[Online; accessed 15-October-2024]}.


\bibitem[{Google}(2024)]%
        {google2024}
\bibfield{author}{\bibinfo{person}{{Google}}.} \bibinfo{year}{2024}\natexlab{}.
\newblock \bibinfo{title}{{Google Health AI}}.
\newblock \bibinfo{howpublished}{\url{https://ai.google/discover/healthai/}}.
\newblock
\newblock
\shownote{[Online; accessed 15-October-2024]}.


\bibitem[Hsieh et~al\mbox{.}(2023)]%
        {hsieh2023what}
\bibfield{author}{\bibinfo{person}{Gary Hsieh}, \bibinfo{person}{Brett~A. Halperin}, \bibinfo{person}{Evan Schmitz}, \bibinfo{person}{Yen~Nee Chew}, {and} \bibinfo{person}{Yuan-Chi Tseng}.} \bibinfo{year}{2023}\natexlab{}.
\newblock \showarticletitle{What is in the Cards: Exploring Uses, Patterns, and Trends in Design Cards}. In \bibinfo{booktitle}{\emph{Proceedings of the 2023 CHI Conference on Human Factors in Computing Systems}} (Hamburg, Germany) \emph{(\bibinfo{series}{CHI '23})}. \bibinfo{publisher}{Association for Computing Machinery}, \bibinfo{address}{New York, NY, USA}, Article \bibinfo{articleno}{765}, \bibinfo{numpages}{18}~pages.
\newblock
\showISBNx{9781450394215}
\urldef\tempurl%
\url{https://doi.org/10.1145/3544548.3580712}
\showDOI{\tempurl}


\bibitem[Jacobs et~al\mbox{.}(2021)]%
        {jacobs2021designing}
\bibfield{author}{\bibinfo{person}{Maia Jacobs}, \bibinfo{person}{Jeffrey He}, \bibinfo{person}{Melanie F.~Pradier}, \bibinfo{person}{Barbara Lam}, \bibinfo{person}{Andrew~C. Ahn}, \bibinfo{person}{Thomas~H. McCoy}, \bibinfo{person}{Roy~H. Perlis}, \bibinfo{person}{Finale Doshi-Velez}, {and} \bibinfo{person}{Krzysztof~Z. Gajos}.} \bibinfo{year}{2021}\natexlab{}.
\newblock \showarticletitle{Designing AI for Trust and Collaboration in Time-Constrained Medical Decisions: A Sociotechnical Lens}. In \bibinfo{booktitle}{\emph{Proceedings of the 2021 CHI Conference on Human Factors in Computing Systems}} (Yokohama, Japan) \emph{(\bibinfo{series}{CHI '21})}. \bibinfo{publisher}{Association for Computing Machinery}, \bibinfo{address}{New York, NY, USA}, Article \bibinfo{articleno}{659}, \bibinfo{numpages}{14}~pages.
\newblock
\showISBNx{9781450380966}
\urldef\tempurl%
\url{https://doi.org/10.1145/3411764.3445385}
\showDOI{\tempurl}


\bibitem[Karusala et~al\mbox{.}(2021)]%
        {karusala2021future}
\bibfield{author}{\bibinfo{person}{Naveena Karusala}, \bibinfo{person}{Azra Ismail}, \bibinfo{person}{Karthik~S Bhat}, \bibinfo{person}{Aakash Gautam}, \bibinfo{person}{Sachin~R Pendse}, \bibinfo{person}{Neha Kumar}, \bibinfo{person}{Richard Anderson}, \bibinfo{person}{Madeline Balaam}, \bibinfo{person}{Shaowen Bardzell}, \bibinfo{person}{Nicola~J Bidwell}, \bibinfo{person}{Melissa Densmore}, \bibinfo{person}{Elizabeth Kaziunas}, \bibinfo{person}{Anne~Marie Piper}, \bibinfo{person}{Noopur Raval}, \bibinfo{person}{Pushpendra Singh}, \bibinfo{person}{Austin Toombs}, \bibinfo{person}{Nervo Verdezoto}, {and} \bibinfo{person}{Ding Wang}.} \bibinfo{year}{2021}\natexlab{}.
\newblock \showarticletitle{The Future of Care Work: Towards a Radical Politics of Care in CSCW Research and Practice}. In \bibinfo{booktitle}{\emph{Companion Publication of the 2021 Conference on Computer Supported Cooperative Work and Social Computing}} (Virtual Event, USA) \emph{(\bibinfo{series}{CSCW '21 Companion})}. \bibinfo{publisher}{Association for Computing Machinery}, \bibinfo{address}{New York, NY, USA}, \bibinfo{pages}{338–342}.
\newblock
\showISBNx{9781450384797}
\urldef\tempurl%
\url{https://doi.org/10.1145/3462204.3481734}
\showDOI{\tempurl}


\bibitem[Kawakami et~al\mbox{.}(2024)]%
        {kawakami2024situate}
\bibfield{author}{\bibinfo{person}{Anna Kawakami}, \bibinfo{person}{Amanda Coston}, \bibinfo{person}{Haiyi Zhu}, \bibinfo{person}{Hoda Heidari}, {and} \bibinfo{person}{Kenneth Holstein}.} \bibinfo{year}{2024}\natexlab{}.
\newblock \showarticletitle{The Situate AI Guidebook: Co-Designing a Toolkit to Support Multi-Stakeholder, Early-stage Deliberations Around Public Sector AI Proposals}. In \bibinfo{booktitle}{\emph{Proceedings of the CHI Conference on Human Factors in Computing Systems}}. \bibinfo{pages}{1--22}.
\newblock


\bibitem[Kim et~al\mbox{.}(2023)]%
        {kim2023organizational}
\bibfield{author}{\bibinfo{person}{Jee~Young Kim}, \bibinfo{person}{William Boag}, \bibinfo{person}{Freya Gulamali}, \bibinfo{person}{Alifia Hasan}, \bibinfo{person}{Henry David~Jeffry Hogg}, \bibinfo{person}{Mark Lifson}, \bibinfo{person}{Deirdre Mulligan}, \bibinfo{person}{Manesh Patel}, \bibinfo{person}{Inioluwa~Deborah Raji}, \bibinfo{person}{Ajai Sehgal}, {et~al\mbox{.}}} \bibinfo{year}{2023}\natexlab{}.
\newblock \showarticletitle{Organizational governance of emerging technologies: AI adoption in healthcare}. In \bibinfo{booktitle}{\emph{proceedings of the 2023 ACM conference on fairness, accountability, and transparency}}. \bibinfo{pages}{1396--1417}.
\newblock


\bibitem[Kurniawan et~al\mbox{.}(2024)]%
        {doi:10.1080/07853890.2024.2302980}
\bibfield{author}{\bibinfo{person}{Moh~Heri Kurniawan}, \bibinfo{person}{Hanny Handiyani}, \bibinfo{person}{Tuti Nuraini}, \bibinfo{person}{Rr~Tutik~Sri Hariyati}, {and} \bibinfo{person}{Sutrisno Sutrisno}.} \bibinfo{year}{2024}\natexlab{}.
\newblock \showarticletitle{A systematic review of artificial intelligence-powered (AI-powered) chatbot intervention for managing chronic illness}.
\newblock \bibinfo{journal}{\emph{Annals of Medicine}} \bibinfo{volume}{56}, \bibinfo{number}{1} (\bibinfo{year}{2024}), \bibinfo{pages}{2302980}.
\newblock


\bibitem[Matheny et~al\mbox{.}(2020)]%
        {matheny2020artificial}
\bibfield{author}{\bibinfo{person}{Michael~E Matheny}, \bibinfo{person}{Danielle Whicher}, {and} \bibinfo{person}{Sonoo~Thadaney Israni}.} \bibinfo{year}{2020}\natexlab{}.
\newblock \showarticletitle{Artificial intelligence in health care: a report from the National Academy of Medicine}.
\newblock \bibinfo{journal}{\emph{Jama}} \bibinfo{volume}{323}, \bibinfo{number}{6} (\bibinfo{year}{2020}), \bibinfo{pages}{509--510}.
\newblock


\bibitem[Montori(2020)]%
        {montori2020we}
\bibfield{author}{\bibinfo{person}{Victor Montori}.} \bibinfo{year}{2020}\natexlab{}.
\newblock \bibinfo{booktitle}{\emph{Why we revolt: a patient revolution for careful and kind care}}.
\newblock \bibinfo{publisher}{Rosetta Books}.
\newblock


\bibitem[Nunes~Vilaza et~al\mbox{.}(2022)]%
        {nunes2022scoping}
\bibfield{author}{\bibinfo{person}{Giovanna Nunes~Vilaza}, \bibinfo{person}{Kevin Doherty}, \bibinfo{person}{Darragh McCashin}, \bibinfo{person}{David Coyle}, \bibinfo{person}{Jakob Bardram}, {and} \bibinfo{person}{Marguerite Barry}.} \bibinfo{year}{2022}\natexlab{}.
\newblock \showarticletitle{A scoping review of ethics across SIGCHI}. In \bibinfo{booktitle}{\emph{Proceedings of the 2022 ACM Designing Interactive Systems Conference}}. \bibinfo{pages}{137--154}.
\newblock


\bibitem[Obermeyer et~al\mbox{.}(2019)]%
        {obermeyer2019racial-bias}
\bibfield{author}{\bibinfo{person}{Ziad Obermeyer}, \bibinfo{person}{Brian Powers}, \bibinfo{person}{Christine Vogeli}, {and} \bibinfo{person}{Sendhil Mullainathan}.} \bibinfo{year}{2019}\natexlab{}.
\newblock \showarticletitle{Dissecting racial bias in an algorithm used to manage the health of populations}.
\newblock \bibinfo{journal}{\emph{Science}} \bibinfo{volume}{366}, \bibinfo{number}{6464} (\bibinfo{year}{2019}), \bibinfo{pages}{447--453}.
\newblock


\bibitem[Organization et~al\mbox{.}(2005)]%
        {world2005mental}
\bibfield{author}{\bibinfo{person}{World~Health Organization} {et~al\mbox{.}}} \bibinfo{year}{2005}\natexlab{}.
\newblock \bibinfo{booktitle}{\emph{Mental Health: Facing the Challenges, Building Solutions. Report From the WHO European Ministerial Conference}}.
\newblock \bibinfo{publisher}{WHO Regional Office Europe}.
\newblock


\bibitem[Osman~Andersen et~al\mbox{.}(2021)]%
        {andersen2021realizing}
\bibfield{author}{\bibinfo{person}{Tariq Osman~Andersen}, \bibinfo{person}{Francisco Nunes}, \bibinfo{person}{Lauren Wilcox}, \bibinfo{person}{Elizabeth Kaziunas}, \bibinfo{person}{Stina Matthiesen}, {and} \bibinfo{person}{Farah Magrabi}.} \bibinfo{year}{2021}\natexlab{}.
\newblock \showarticletitle{Realizing AI in Healthcare: Challenges Appearing in the Wild}. In \bibinfo{booktitle}{\emph{Extended Abstracts of the 2021 CHI Conference on Human Factors in Computing Systems}} (Yokohama, Japan) \emph{(\bibinfo{series}{CHI EA '21})}. \bibinfo{publisher}{Association for Computing Machinery}, \bibinfo{address}{New York, NY, USA}, Article \bibinfo{articleno}{108}, \bibinfo{numpages}{5}~pages.
\newblock
\showISBNx{9781450380959}
\urldef\tempurl%
\url{https://doi.org/10.1145/3411763.3441347}
\showDOI{\tempurl}


\bibitem[Sadek et~al\mbox{.}(2024a)]%
        {sadek2024guidelines}
\bibfield{author}{\bibinfo{person}{Malak Sadek}, \bibinfo{person}{Marios Constantinides}, \bibinfo{person}{Daniele Quercia}, {and} \bibinfo{person}{C{\'e}line Mougenot}.} \bibinfo{year}{2024}\natexlab{a}.
\newblock \showarticletitle{Guidelines for Integrating Value Sensitive Design in Responsible AI Toolkits}. In \bibinfo{booktitle}{\emph{Proceedings of the CHI Conference on Human Factors in Computing Systems}}. \bibinfo{pages}{1--20}.
\newblock


\bibitem[Sadek et~al\mbox{.}(2024b)]%
        {sadek2024challenges}
\bibfield{author}{\bibinfo{person}{Malak Sadek}, \bibinfo{person}{Emma Kallina}, \bibinfo{person}{Thomas Bohn{\'e}}, \bibinfo{person}{C{\'e}line Mougenot}, \bibinfo{person}{Rafael~A Calvo}, {and} \bibinfo{person}{Stephen Cave}.} \bibinfo{year}{2024}\natexlab{b}.
\newblock \showarticletitle{Challenges of responsible AI in practice: scoping review and recommended actions}.
\newblock \bibinfo{journal}{\emph{AI \& SOCIETY}} (\bibinfo{year}{2024}), \bibinfo{pages}{1--17}.
\newblock


\bibitem[Sch{\"u}ll(2016)]%
        {schull2016data}
\bibfield{author}{\bibinfo{person}{Natasha~Dow Sch{\"u}ll}.} \bibinfo{year}{2016}\natexlab{}.
\newblock \showarticletitle{Data for life: Wearable technology and the design of self-care}.
\newblock \bibinfo{journal}{\emph{BioSocieties}}  \bibinfo{volume}{11} (\bibinfo{year}{2016}), \bibinfo{pages}{317--333}.
\newblock


\bibitem[Shaw et~al\mbox{.}(2019)]%
        {shaw2019artificial}
\bibfield{author}{\bibinfo{person}{James Shaw}, \bibinfo{person}{Frank Rudzicz}, \bibinfo{person}{Trevor Jamieson}, {and} \bibinfo{person}{Avi Goldfarb}.} \bibinfo{year}{2019}\natexlab{}.
\newblock \showarticletitle{Artificial intelligence and the implementation challenge}.
\newblock \bibinfo{journal}{\emph{Journal of medical Internet research}} \bibinfo{volume}{21}, \bibinfo{number}{7} (\bibinfo{year}{2019}), \bibinfo{pages}{e13659}.
\newblock


\bibitem[Slattery et~al\mbox{.}(2024)]%
        {slattery2024ai}
\bibfield{author}{\bibinfo{person}{Peter Slattery}, \bibinfo{person}{Alexander~K Saeri}, \bibinfo{person}{Emily~AC Grundy}, \bibinfo{person}{Jess Graham}, \bibinfo{person}{Michael Noetel}, \bibinfo{person}{Risto Uuk}, \bibinfo{person}{James Dao}, \bibinfo{person}{Soroush Pour}, \bibinfo{person}{Stephen Casper}, {and} \bibinfo{person}{Neil Thompson}.} \bibinfo{year}{2024}\natexlab{}.
\newblock \showarticletitle{The AI Risk Repository: A Comprehensive Meta-Review, Database, and Taxonomy of Risks From Artificial Intelligence}.
\newblock \bibinfo{journal}{\emph{arXiv preprint arXiv:2408.12622}} (\bibinfo{year}{2024}).
\newblock


\bibitem[van Kolfschooten and van Oirschot(2024)]%
        {van2024eu}
\bibfield{author}{\bibinfo{person}{Hannah van Kolfschooten} {and} \bibinfo{person}{Janneke van Oirschot}.} \bibinfo{year}{2024}\natexlab{}.
\newblock \showarticletitle{The EU Artificial Intelligence Act (2024): Implications for healthcare}.
\newblock \bibinfo{journal}{\emph{Health Policy}}  \bibinfo{volume}{149} (\bibinfo{year}{2024}), \bibinfo{pages}{105152}.
\newblock


\bibitem[Vorvoreanu(2023)]%
        {vorvoreanu2023create}
\bibfield{author}{\bibinfo{person}{Mihaela Vorvoreanu}.} \bibinfo{year}{2023}\natexlab{}.
\newblock \showarticletitle{Create Effective and Responsible AI User Experiences with The Human-AI Experience (HAX) Toolkit}. In \bibinfo{booktitle}{\emph{Extended Abstracts of the 2023 CHI Conference on Human Factors in Computing Systems}}. \bibinfo{pages}{1--2}.
\newblock


\bibitem[Waefler and Schmid(2020)]%
        {waefler2020explainability}
\bibfield{author}{\bibinfo{person}{Toni Waefler} {and} \bibinfo{person}{Ute Schmid}.} \bibinfo{year}{2020}\natexlab{}.
\newblock \showarticletitle{Explainability is not enough: requirements for human-AI-partnership in complex socio-technical systems}.
\newblock  (\bibinfo{year}{2020}).
\newblock


\bibitem[Wong et~al\mbox{.}(2023)]%
        {wong2023seeing}
\bibfield{author}{\bibinfo{person}{Richmond~Y Wong}, \bibinfo{person}{Michael~A Madaio}, {and} \bibinfo{person}{Nick Merrill}.} \bibinfo{year}{2023}\natexlab{}.
\newblock \showarticletitle{Seeing like a toolkit: How toolkits envision the work of AI ethics}.
\newblock \bibinfo{journal}{\emph{Proceedings of the ACM on Human-Computer Interaction}} \bibinfo{volume}{7}, \bibinfo{number}{CSCW1} (\bibinfo{year}{2023}), \bibinfo{pages}{1--27}.
\newblock


\bibitem[Yang et~al\mbox{.}(2019)]%
        {yang2019unremarkable}
\bibfield{author}{\bibinfo{person}{Qian Yang}, \bibinfo{person}{Aaron Steinfeld}, {and} \bibinfo{person}{John Zimmerman}.} \bibinfo{year}{2019}\natexlab{}.
\newblock \showarticletitle{Unremarkable AI: Fitting Intelligent Decision Support into Critical, Clinical Decision-Making Processes}. In \bibinfo{booktitle}{\emph{Proceedings of the 2019 CHI Conference on Human Factors in Computing Systems -- CHI '19}}. \bibinfo{pages}{1--11}.
\newblock


\bibitem[Yildirim et~al\mbox{.}(2023)]%
        {yildirim2023creating}
\bibfield{author}{\bibinfo{person}{Nur Yildirim}, \bibinfo{person}{Changhoon Oh}, \bibinfo{person}{Deniz Sayar}, \bibinfo{person}{Kayla Brand}, \bibinfo{person}{Supritha Challa}, \bibinfo{person}{Violet Turri}, \bibinfo{person}{Nina Crosby~Walton}, \bibinfo{person}{Anna~Elise Wong}, \bibinfo{person}{Jodi Forlizzi}, \bibinfo{person}{James McCann}, {et~al\mbox{.}}} \bibinfo{year}{2023}\natexlab{}.
\newblock \showarticletitle{Creating design resources to scaffold the ideation of AI concepts}. In \bibinfo{booktitle}{\emph{Proceedings of the 2023 ACM Designing Interactive Systems Conference}}. \bibinfo{pages}{2326--2346}.
\newblock


\bibitem[Zaj{\k{a}}c et~al\mbox{.}(2023)]%
        {zajkac2023clinician}
\bibfield{author}{\bibinfo{person}{Hubert~D Zaj{\k{a}}c}, \bibinfo{person}{Dana Li}, \bibinfo{person}{Xiang Dai}, \bibinfo{person}{Jonathan~F Carlsen}, \bibinfo{person}{Finn Kensing}, {and} \bibinfo{person}{Tariq~O Andersen}.} \bibinfo{year}{2023}\natexlab{}.
\newblock \showarticletitle{Clinician-facing AI in the Wild: Taking Stock of the Sociotechnical Challenges and Opportunities for HCI}.
\newblock \bibinfo{journal}{\emph{ACM Transactions on Computer-Human Interaction}} \bibinfo{volume}{30}, \bibinfo{number}{2} (\bibinfo{year}{2023}), \bibinfo{pages}{1--39}.
\newblock


\bibitem[Ziebland et~al\mbox{.}(2021)]%
        {ziebland2021power}
\bibfield{author}{\bibinfo{person}{Sue Ziebland}, \bibinfo{person}{Emma Hyde}, {and} \bibinfo{person}{John Powell}.} \bibinfo{year}{2021}\natexlab{}.
\newblock \showarticletitle{Power, paradox and pessimism: on the unintended consequences of digital health technologies in primary care}.
\newblock \bibinfo{journal}{\emph{Social Science \& Medicine}}  \bibinfo{volume}{289} (\bibinfo{year}{2021}), \bibinfo{pages}{114419}.
\newblock


\end{thebibliography}

\end{document}